\documentclass[manuscript]{aastex}
\shorttitle{Fast Chi-Squared Period Search}
\shortauthors{Palmer}
\slugcomment{Submitted to ApJ}
\def\eg{{e.g}}	
\def\gae{{_{\sim}^>}}
\def\lae{{_{\sim}^<}}

\def\chisquared{{$\chi^2$}}
\def\dc2{{$\Delta \chi^2$}}
\def\dca2{{$\Delta \chi_{adj}^2$}}
\def\fc2{{${\rm F}\chi^2$}}
\def\mfc2{{{\rm F}\chi^2}}

\newcommand{\leftsub}[2]{{\vphantom{#2}}_{#1}{#2}}

\begin{document}

\title{A Fast Chi-squared Technique For Period Search of Irregularly Sampled Data.}
\author{David M. Palmer}
\affil{Los Alamos National Laboratory, B244, Los Alamos, NM 87545}
\email{palmer@lanl.gov}

\begin{abstract}

A new, computationally- and statistically-efficient algorithm,
the Fast $\chi^2$ algorithm (\fc2), can find a periodic signal
with harmonic content in irregularly-sampled data with non-uniform
errors.
The algorithm calculates the minimized $\chi^2$ as a function of
frequency at the desired number of harmonics, using Fast Fourier Transforms
to provide $O ( N \log N )$ performance.  The code
for a reference implementation is provided.
\end{abstract}


\maketitle

\section{Introduction}

A common problem, in astronomy and in other fields, is to detect the presence
and characteristics of an unknown periodic signal in irregularly-spaced
data.  Throughout this paper I will use examples and language drawn
from the study of periodic variable stars, but the techniques can be
applied to many other situations.

Generally, detecting periodicity is achieved by comparing 
the measurements phased at each of a set of trial frequencies
to a model periodic 
phase function, $\Phi(\phi)$, and selecting the frequency that yields a
high value for a quality function.
The commonly-used techniques vary in the form and parameterization
of $\Phi$, the evaluation of the fit quality between model and data, the set of
frequencies searched, and
the methods used for computational efficiency.

The Lomb algorithm \citep{1976Ap&SS..39..447L}, for example, uses a sinusoid plus constant 
for its model function.  The quality
is the amplitude of the unweighted least-squares fit at the trial frequency.
A simple implementation takes $O ( n N )$ computations, where $n$ is the number of 
measurements and $N$ is the number of frequencies searched. 
\citet{1989ApJ...338..277P}
present an implementation that uses the Fast Fourier Transform (FFT)
to give
$O(N log N)$ performance.  Spurious detections can be produced at frequencies where the 
sample times have significant periodicity, power from harmonics above the 
fundamental sine wave is lost, and the algorithm is statistically inefficient
in the sense that it ignores point-to-point variation in the measurement error.

\citet{Irwin} use a Lomb algorithm to find a candidate period, then test the
measurements' \chisquared reduction between a constant value and a model periodic lightcurve
obtained by folding the measurements at that period and applying smoothing.

Phase Dispersion Minimization \citep{1978ApJ...224..953S} divides a cycle 
into (possibly overlapping)
phase bins.  The quality is calculated from the \chisquared agreement among those data points
that fall into each bin at the trial frequency.  This also has $O ( n N )$ performance.
The size and number of the bins, and the time of $\phi = 0$, are choices that can
affect the detection probability of a particular signal.

The Fast \chisquared (hereafter \fc2) technique presented here uses a Fourier series
truncated at harmonic $H$:
\begin{equation}
\Phi_{H}(\{A_{0\ldots 2H},f\},t) = 
A_0 + \sum_{h=1\ldots H} A_{2h-1} \sin(h 2 \pi f t) + A_{2h} \cos(h 2 \pi f t)
\label{eqtrunc}
\end{equation}
for the model periodic function.  The fit quality is the \chisquared
of all data, jointly minimized over the Fourier coefficients, $A_{0{\ldots}2H}$,
and the frequency, $f$.
\citet{baluev08} investigates the statistics of these fits.
The set of frequencies searched can
have arbitrary density and range.  The computational
complexity of this implementation is $O(H N \log {H N} )$.  The optimal choice
of $H$ depends on the (generally-unknown) true shape of the periodic signal, but evaluations
with multiple $H$ values can share computational stages, resulting in
determinations at $H^\prime = 1{\ldots}H$ at only a slight premium.
The fit follows standard \chisquared statistics and makes efficient use of
measurement errors.

\section{Description}

\chisquared minimization is the standard method to fit a data set with
unequal, Gaussian, measurement errors.
(Other Maximum Likelihood methods
can handle more general error distributions,
but are beyond the scope of this paper.)

A hypothesis may be expressed as a model $M({\bf P}, t)$,
where ${\bf P}$ is the set of unknown model parameters
and $t$ is an independent variable.
This is compared to measurements $x_i$, with standard errors
$\sigma_i$, made at $t_i$.  The comparison is made by finding the ${\bf P}$
that minimizes 
\begin{equation}
\chi^2({\bf  P}) = \sum_{i=1 \ldots n} \frac{(M({\bf  P}, t_i) - x_i)^2}{\sigma_i^2}.
\label{eqchigen}
\end{equation}
(In this paper $t$ is a scalar value which, for the variable star case, represents time.
However, the extension of this
technique to spatial or other dimensions, 
to higher dimension such as the $(u,v)$ baselines of interferometry,
and to non-scalar measurements, ${\bf x_i}$, is straightforward.)

If this minimum \chisquared is significantly below that for a Null Hypothesis, then
this is evidence for the model, and for the value of ${\bf P}$ at the minimum.
The extent by which the model decreases the \chisquared compared to the Null hypotheses, 
$\Delta \chi^2 = \chi^2_0 - \chi^2({\bf  P})$, is known as the Minimization Index and, 
if the Null Hypothesis is true and other conditions apply, has a \chisquared probability 
distribution with the same number of Degrees Of Freedom as the model.

Many models of interest are linear in some parameters, and non-linear in others.
These models can be factored into the linear parameters, ${\bf A}$, and
a set of functions ${\bf M}({\bf Z}, t)$ of the non-linear parameters, ${\bf Z}$, so that
\begin{equation}
M(\{{\bf A},{\bf Z}\}, t) = \sum_{i} {\bf A}_i {\bf M}_{i}({\bf Z}, t)
\end{equation}

For any given ${\bf Z}$, the \chisquared minimized with respect to ${\bf A}$
can be rapidly found in closed form by linear regression.
Finding the global minimum of \chisquared thus reduces to searching the
non-linear parameter space ${\bf Z}$.

The truncated Fourier series function in equation \ref{eqtrunc}
is linear in the coefficients $A_{0\ldots 2H}$, and non-linear in ${\bf Z} = f$,
with component functions
\begin{eqnarray}
{\bf M}_{0}(f, t) &=& 1 \\
{\bf M}_{2h-1}(f, t) &=& \sin h 2 \pi f t \\
{\bf M}_{2h}(f, t) &=& \cos h 2 \pi f t 
\end{eqnarray}
Therefore, the minimum \chisquared can be quickly calculated at any chosen frequency.
Complete, dense coverage of the frequency range of interest is required to find
the overall minimum---the orthogonality of the Fourier basis tends to make the minima
very narrow.

Proceeding in the usual manner for linear regression (see \eg,
{\it{Numerical Recipes}}
\citep{NumRec}, 
\S 15.4 for a review) we produce the $\alpha$ matrix and $\beta$ vector
used in the `normal equations'.  The values of $\alpha$ and $\beta$ depend
on the frequency $f$.

The $\alpha$ matrix comes from the cross terms of the model components,
as a weighted sum over the data points:
\begin{equation}
 ^{\scriptscriptstyle f}\alpha_{jk}= \sum_{i=1 \ldots n}
				\frac{{\bf M}_j(f, t_i) {\bf M}_k(f, t_i)}
					{\sigma_i^2} 
\end{equation}

The $\beta$ vector is the weighted sum of the product of the data
and the model components: 
\begin{equation}
 ^{\scriptscriptstyle f}{\bf \beta}_{j} = \sum_{i=1 \ldots n}
				\frac{{\bf M}_j(f, t_i) x_i} 
					{\sigma_i^2} 
\end{equation}

A frequency-independent scalar, $X$, which is the \chisquared corresponding to
a $constant = 0$ model:
\begin{equation}
{\rm X} = \sum_i {x_i^2}/{\sigma_i^2}
\end{equation}
is also used in the calculation of \chisquared.

The minimization of equation \ref{eqchigen} over the linear ${\bf A}$
parameters can be shown to give a minimum  \chisquared subject to $f$ of
\begin{equation}
^{\scriptscriptstyle f}\chi^2_{min} = {\rm X} - \sum_j \sum_k {} ^{\scriptscriptstyle f}{\bf \beta}_{j}{}
 (^{\scriptscriptstyle f}\alpha^{-1})_{jk}{} ^{\scriptscriptstyle f}{\bf \beta}_{k}
\label{eqchix}
\end{equation}

The Fourier coefficients corresponding to this minimum are
\begin{equation}
^{\scriptscriptstyle f}{\bf A}_j = \sum_k (^{\scriptscriptstyle f}\alpha^{-1})_{jk}{} ^{\scriptscriptstyle f}{\bf \beta}_{k}
\label{eqcoeffs}
\end{equation}

We can use the trigonometric product relations:
\begin{eqnarray}
\sin(a)\sin(b) &= {\scriptstyle \frac12} (\cos(a-b) - \cos(a+b)) \\
\sin(a)\cos(b) &= {\scriptstyle \frac12} (\sin(a+b) + \sin(a-b)) \\
\cos(a)\sin(b) &= {\scriptstyle \frac12}  (\sin(a+b) - \sin(a-b)) \\
\cos(a)\cos(b) &= {\scriptstyle \frac12} (\cos(a-b) + \cos(a+b))
\end{eqnarray}
to reduce the cross-terms of sines and cosines in $\alpha$ to
\begin{eqnarray}
\label{eqalphafirst}
^{\scriptscriptstyle f}\alpha_{2h-1,2h'-1}&= {\scriptstyle \frac12} ( C'((h-h')f) - C'((h+h')f)) \\
^{\scriptscriptstyle f}\alpha_{2h-1,2h'} &= {\scriptstyle \frac12}  (S'((h+h')f) + S'((h-h')f))\\
^{\scriptscriptstyle f}\alpha_{2h,2h'-1} &= {\scriptstyle \frac12}  (S'((h+h')f) - S'((h-h')f))  \\
^{\scriptscriptstyle f}\alpha_{2h,2h'} &= {\scriptstyle \frac12} (C'((h-h')f) + C'((h+h')f))
\label{eqalpha}
\end{eqnarray}
while the terms for $\beta$ are
\begin{eqnarray}
^{\scriptscriptstyle f}{\bf \beta}_{2h-1} = S(hf) \\
^{\scriptscriptstyle f}{\bf \beta}_{2h} = C(hf)
\end{eqnarray}
where
\begin{equation}
\begin{array}{rcccl}
C(f) &=& \sum{\cos( 2 \pi f t_i) x_i}/{\sigma_i^2} &=&{\rm re}\ P(f)\\
S(f) &=& \sum{\sin( 2 \pi f t_i) x_i}/{\sigma_i^2}  &=& {\rm im}\ P(f)\\
C'(f) &=& \sum{\cos( 2 \pi f t_i)}/{\sigma_i^2}  &=& {\rm re}\ Q(f)\\
S'(f) &=& \sum{\sin( 2 \pi f t_i)}/{\sigma_i^2}  &=& {\rm  im}\ Q(f)
\end{array}
\end{equation}

The functions $C, C', S,$ and $S'$ are the cosine and sine parts of the Fourier transforms
of the weighted data and the weights:
\begin{equation}
\begin{array}{rclcrcl}
p(t)&=&\sum_i \delta(t - t_i) \frac{x_i}{\sigma_i^2} &\hspace{1cm}&P(f) = \int p(t) e^{i2 \pi f t} dt\\
q(t)&=&\sum_i \delta(t - t_i) \frac{1}{\sigma_i^2} &&Q(f) = \int q(t) e^{i2 \pi f t} dt
\label{eqPQ}
\end{array}
\end{equation}

The function $p(t)$ is the weighted data, and $q(t)$ is the weight, as a function of time.  
These are both zero at times when there is no data, and Dirac delta functions at the time of each measurement.
$P$ and $Q$ can be efficiently computed by the application of the FFT to $p$ and $q$.

This FFT calculation of $P$ and $Q$, their use in construction of $\alpha$ and $\beta$ for
a set of frequencies, and the \chisquared minimization by equation \ref{eqchix} for each of those
frequencies, are the components of the \fc2 algorithm.

\section{Implementation}

The steps in implementing a \fc2 search are:
a) Choosing the search space
b) Generating the Fourier Transforms
c) Calculating the normal equations at each frequency
d) Finding the minimum
e) Interpreting the result

{\it a) Choosing the search space}

The frequency range of interest, the number of harmonics,
and the density of coverage
may be chosen based on physical expectations,
measurement characteristics, processing limitations, or other considerations.

The maximum frequency searched may be where the exposure times 
of the observations are a large fraction of the period of the highest harmonic, 
or some other upper bound
placed by experience or the physics of the source.  At low frequencies, if
there are only a few cycles or less of the fundamental over the span of all observations,
a large \chisquared decrease may be evidence of variability but not necessarily of periodicity.

It is not necessary that the fitting function accurately represent all aspects of
the physical process.   Sharp features in the actual phase function
may require high harmonics to reproduce as a Fourier sum,
but can be detected with adequate sensitivity using a small number of harmonics.
Details of the phase function 
(\eg, the behavior on ingress and egress of an eclipsing binary) 
may be determined by other techniques once a candidate frequency has been found.

The density of the search--how closely the trial frequencies are spaced--affects
the sensitivity as well.  For a simple Fourier analysis of the fundamental, a spacing
of one cycle over the span of observations can cause a reduction in amplitude
to $1/\sqrt(2)$ if the true frequency is intermediate between
two trial frequencies.  For an analysis to harmonic $H$, the spacing of trial fundamental
frequencies must be correspondingly tighter.  The maximum sensitivity loss depends
on the harmonic content of the signal, which is generally not {\it a priori} known, but a
spacing much looser than $1/H$ cycles will typically lose the advantage
of going to harmonic $H$, while a much tighter spacing will consume
resources that might be better employed with a search to $H+1$, depending on the
characteristics of the source.

If $\Delta T$ is the timespan of all observations and $\delta t$ is the fastest
timescale of interest,
then a reasonable maximum frequency and
spacing for the fundamental would be $f_{max} \gae 1/(2 H \delta t)$ 
and $\delta f \lae 1/(H \Delta T)$.

{\it b) Generating the Fourier Transforms}

The calculation of $^{\scriptscriptstyle f}\chi^2_{min}$ requires evaluation of the Fourier functions
$P$, at $\{0,f\ldots H f\}$, and
$Q$, at $\{0,f\ldots 2 H f\}$.  

The real-to-complex FFT, as typically implemented, takes as input $2 \mathcal{N}$ real
data points from uniformly-spaced discrete times
$t=\{0,\delta t,2\delta t\ldots( 2 {\mathcal{N}}-1) \delta t\}$.
If the data were not sampled at those exact
times, they are `gridded' 
(\eg, by interpolation, or by nearest-neighbor sampling)
to estimate what the data values at those times would have been.
The output of the FFT is the ${\mathcal{N}} + 1$ complex
Fourier components
corresponding to frequencies $f = \{0,\delta f,2\delta f\ldots {\mathcal{N}} \delta f\}$, where 
$\delta f = 1/(2 {\mathcal{N}} \delta t)$.

To provide the frequency range and density required for 
the \fc2 method, the weighted-data and the weights are placed
in sparse, zero-padded arrays, with $\delta t$ bin width,
that cover at least $2H$ times the observed
time period.  This produces discrete
functions (using $\widehat{\rm hat}$ symbols to indicate discrete quantities)
\begin{eqnarray}
\hat{p}[\hat{t}] &= \sum_i \hat{\delta}[\hat{t},\hat{t}_i] \frac{x_i - \bar{x}}{\sigma_i^2} \\
\hat{q}[\hat{t}] &= \sum_i \hat{\delta}[\hat{t},\hat{t}_i] \frac{1}{\sigma_i^2}
\end{eqnarray}
In this notation, hatted times are integer indices offset by a starting epoch,
$T_0 \le {\rm all}\ t_i$,
scaled by $\delta t$, and
rounded down to the lower integer:
$\hat{t} = {\rm floor}((t - T_0)/\delta t)$; and $\hat{\delta}$ is the Kronecker delta:
\begin{equation}
\hat{\delta}[m,n]  = \left\{ {\begin{array}{cc} 
1 & \hspace{1cm} m = n \\
0 & \hspace{1cm} m \not= n \\
\end{array} } \right.
\end{equation}

For numerical reasons, the measurements are adjusted by the mean
value,
\begin{equation}
 \bar{x} = \frac{\sum_j \frac{x_j}{\sigma_j^2}}{\sum_j \frac{1}{\sigma_j^2}}
\end{equation}
and the value of $A_0$ found by the algorithm should have $\bar{x}$ added back
to find the true mean value of the source.


When implemented, these discrete functions can be represented as arrays
with indices $[0 \ldots 2{\mathcal N} - 1]$.
To search an acceptable density of frequencies, you should choose
$({ \mathcal {N} }/H)\delta t  \gae \Delta T$.
Most practical implementations of the FFT place requirements on
${\mathcal N}$ for the sake of efficiency, such as being a power of 2 or having 
only small prime factors.


The real arrays $\hat{p}[\hat{t}] $ and $\hat{q}[\hat{t}]$ are then passed to an
FFT routine to get the complex arrays $\hat{P}[\hat{f}]$ and $\hat{Q}[\hat{f}]$.
The discrete frequency indices $\hat{f} = 0 \ldots {\mathcal N} - 1$ correspond
to frequencies $f = \hat{f} \delta f$.  (Many FFT implementations
use the imaginary part of the $\hat{f} = 0$ element to store the cosine component 
at the Nyquist frequency $f_{Nyquist} = {\mathcal N}\delta f$.)

{\it c) Calculating the normal equations at each frequency}

Equations \ref{eqalphafirst}--\ref{eqPQ} describe how to construct $\alpha$ and $\beta$.
The $\alpha$ matrix at a given $\hat{f}$ is based on the terms of $\hat{Q}$ at indices
$\{0,\hat{f},2\hat{f}\ldots 2 H \hat{f}\}$.
The $\beta$ vector is based on the terms of $\hat{P}$ at $\{0,\hat{f},2\hat{f}\ldots H \hat{f}\}$.

\def\Qi{{\leftsub{\mathcal I}Q}}
\def\Qr{{\leftsub{\mathcal R}Q}}
\def\Pi{{\leftsub{\mathcal I}P}}
\def\Pr{{\leftsub{\mathcal R}P}}

Streamlining the notation so that  $\hat{Q}[n \hat{f}] = \Qr_n + i~\Qi_n$ and $\hat{P}[n \hat{f}] = \Pr_n + i~\Pi_n$,
the $H=2$ case can be written:

\begin{equation}
{^{\scriptscriptstyle \hat{f}}}\alpha = \frac12 \left( 
\begin{array}{c cc cc}
2~\Qr_0
	&2~\Qi_1&2~\Qr_1
		&2~\Qi_2&2~\Qr_2 \\
2~\Qi_1
	& \Qr_0 - \Qr_2 & \Qi_2
		& \Qr_1 - \Qr_3 & \Qi_3 + \Qi_1\\
2~\Qr_1
	& \Qi_2 &  \Qr_0 + \Qr_2
		& \Qi_3 - \Qi_1 &  \Qr_1+ \Qr_3 \\
2~\Qi_2
	& Qr_1 - \Qr_3 & \Qi_3 - \Qi_1
		& \Qr_0 - \Qr_4 & \Qi_4 \\
2~\Qr_2
	& \Qi_3 + \Qi_1 &  \Qr_1 + \Qr_3
		& \Qi_4 &  \Qr_0 + \Qr_4 \\
\end{array}
\right)
\end{equation}
\begin{equation}
^{\scriptscriptstyle \hat{f}}\beta =  \left( 
\begin{array}{c}
\Pr_0 \\
\Pi_1\\
\Pr_1\\
\Pi_2\\
\Pr_2\\
\end{array}
\right)
\end{equation}


The minimum ${^{\scriptscriptstyle \hat{f}}}\chi^2$ at each frequency is
less than or equal to that for the constant value Null Hypothesis:
\begin{eqnarray}
\chi^2_0 &=&  \sum_i \frac{(x_i - \bar{x})^2}{\sigma_j^2} \\
{^{\scriptscriptstyle \hat{f}}}\chi^2 &=& \chi^2_0 - 
{\sum_{j,k}}
{^{\scriptscriptstyle \hat{f}}}{\bf \beta}_j
({^{\scriptscriptstyle \hat{f}}}\alpha^{-1})_{jk}
{^{\scriptscriptstyle \hat{f}}}{\bf \beta}_k\\
&=& \chi^2_0 - 
^{\scriptscriptstyle \hat{f}}{\Delta \chi^2}
\end{eqnarray}

The values of the Fourier Coefficients, $^{\scriptscriptstyle \hat{f}}{\bf A}_j = {\sum_{k}}
({^{\scriptscriptstyle \hat{f}}}\alpha^{-1})_{jk}
{^{\scriptscriptstyle \hat{f}}}{\bf \beta}_k$,
are not required for finding the minimum.  However, they will typically be calculated as an 
intermediate result and may be used for further analysis.

{\it e) Interpreting the result}

The value of $\hat{f}$ with the largest value of
$
^{\scriptscriptstyle \hat{f}}{\Delta \chi^2} =
{\sum_{j,k}}
{^{\scriptscriptstyle \hat{f}}}{\bf \beta}_j
({^{\scriptscriptstyle \hat{f}}}\alpha^{-1})_{jk}
{^{\scriptscriptstyle \hat{f}}}{\bf \beta}_k$
(and thus the lowest \chisquared)
provides the best fit among the searched frequencies.

The $^{\scriptscriptstyle \hat{f}}{\Delta \chi^2}$ value
tells how much better the model at that frequency fits the data than the
constant-value Null Hypothesis does.
 $^{\scriptscriptstyle \hat{f}}{\Delta \chi^2}$ must be compared to what is expected
by chance, given the number of additional free parameters in the model
($2H$)
and the number of trials (representing independent frequencies searched).

The number of  independent frequencies searched can be
made arbitrarily large by decreasing the gridding interval,
and so any given \chisquared improvement can in theory be diluted away
to insignificance.
However, the number of trials is only unbounded towards higher
frequencies.  If you search starting at low frequencies, the
number of trials `so far' can be treated as being proportional to $f$.
In that case, the value to be minimized is 
\begin{equation}
p(f) = \hat{f} {\mathcal P}_{2H,\ge \Delta \chi^{2}}(^{\scriptscriptstyle \hat{f}}{\Delta \chi^2})
\end{equation}
where ${\mathcal P}_{2H,\ge \Delta \chi^{2}}$
is the cumulative \chisquared distribution with $2H$ degrees of 
freedom.  $p(f)$ is proportional to the probability, given the Null Hypothesis,
of finding such a large decrease in \chisquared by chance at a
frequency $f$ or below. 
 
The use of $f$ to adjust ${\mathcal P}$ is straightforward from
a frequentist perspective.  From a Bayesian perspective it corresponds
to a prior assumption that the distribution of $log(f)$ is uniform
(which is equivalent to $log(period)$ being uniform) over the search interval.
A different adjustment could be made if a different Bayesian prior were desired.

Because the true minimum might fall between two adjacent frequency bins,
and because the gridding of the data causes some sensitivity loss,
frequencies in the vicinity of the minimum, and in the vicinities of
other frequencies that have local minima that are almost as good, 
should be searched more finely.  These searches should use 
the ungridded data to directly calculate a locally-optimum
$^{\scriptscriptstyle f}{\Delta \chi^2}$ 
and adjustments.

Multiple candidate frequencies can be extracted and examined
to see if they can reject the Null Hypothesis using other statistics
in combination with  $^{\scriptscriptstyle f}{\Delta \chi^2}$ .
For example, in a search 
for stars with transiting planets a particular shape of light
curve is expected.  Finding an otherwise marginal 
$^{\scriptscriptstyle f}{\Delta \chi^2}$ 
in combination with this light curve shape
would be a convincing detection.
\fc2 can quickly find all candidate frequencies
with marginal or better $^{\scriptscriptstyle f}{\Delta \chi^2}$ .

The reduced-\chisquared, the ratio of the minimized \chisquared to the degrees
of freedom,  is a useful
measure of how well the data fits the model.  However,
even the correct frequency can produce a poor reduced-\chisquared
under several circumstances.  The source may have a light curve that
has sharp features, or is otherwise poorly-described by an
$H$-harmonic Fourier series.  The source may have
`noisy' variations in addition to periodic behavior.  There may be 
multiple frequencies involved, as in Blazhko effect RR Lyraes,
overtone Cepheids, or eclipsing binaries where one of the
components is itself variable.  In all these cases, the detection
of a period can provide a starting point for further analysis of the source's behavior.

Because the source may be noisier (compared to the model) 
than the measurement error, an adjusted \dc2 may be defined
$$\Delta \chi^2_{adj} = \frac{ \chi^2_0 - \chi^2({\bf  P})}{\Delta \chi^2_{best} / N_{dof}}$$
For a noisy source, this will allow significance calculations to be based
on the noise of the source, rather than the measurement error, preventing
false positives due to fitting the source noise.  For
truly constant sources, and for sources that are accurately described
by the harmonic fit, the measurement error will dominate.
This is an improvement over the traditional method of using the standard deviation
of the data as a surrogate for measurement error because it continues to incorporate
the known instrumental characteristics and because it does not overstate the source noise
of a well-behaved variable source.  
For any given fit, the period with the best \dc2 will also have the best \dca2.

The best-fit frequency is not necessarily the `right answer'.
There are several effects that can produce \chisquared minima at
frequencies that are not the frequency of the physical system being
studied.  Some examples of this are shown in \S 4.
For example, a mutual-eclipsing binary of two similar stars may
be better fit at $2f_{orbit}$ than at $f_{orbit}$ for a given $H$.

If the set of sampling times has a strong periodic component, then
this can produce `aliasing' against periodic or long-term
variations.  Irregular sampling improves the behavior
of \chisquared-based period searches.
For alias peaks to be strong, there must be some frequency,
$f_{sample}$ for which a large fraction of the sampling
times are clustered in phase to within a small fraction
of the true source variation timescale.  For many astronomical
datasets
taken from a single location only at night, this is the case
with $f_{sample} = 1/{\rm day}$ or $1/{\rm sidereal day}$.  Depending
on the observation strategy (\eg, observing each field as it transits
the meridian 
{\it vs}. observing several times a night) the clustering in phase can be
tighter or looser, and thus produce greater or lesser aliasing at short
periods.

If there is long-term variation in the source, then there
may be alias peaks near $f_{sample}$ and its harmonics, even if the
long-term
variation is aperiodic.  If the source combines a high-frequency
periodicity with a higher-amplitude long-term aperiodic variation,
then the alias peaks can provide the lowest \chisquared.
To detect the short period variation, the long-term variation
may be removed with, \eg, a polynomial fit (although, as discussed in \S 4,
this will not necessarily result in an improvement).  An initial
run of the \fc2 algorithm with $\delta t = 1/f_{sample}$
can be used to quickly test whether the long-term variations
are themselves periodic.

One advantage of \chisquared methods over Lomb
is that they do not produce alias peaks near $f_{sample}$
if the source is constant.  Near $f_{sample}$, the limited
phase coverage of the samples provides only
weak constraints on the amplitude of the Fourier components.
Statistical fluctuations can produce large amplitudes
(and thus high Lomb values),
but the correspondingly large standard error on the
amplitude ensures that such fits do not yield a large improvement in \chisquared.

\section{Reference Implementation And Examples}

An implementation of the algorithm, written in C,  is available at the author's
website\footnote{\url{http://public.lanl.gov/palmer/fastchi.html}}.
For maximum portability and clarity, it uses the Gnu Scientific Library 
 \citep[GSL, ][]{GSL}
and includes a driver interface that reads the input data as ASCII files.

The processing speed might be improved by using a different FFT package, such as FFTW \citep{FFTW05},
by stripping the GSL and CBLAS layers from the core BLAS routines, and by using 
binary I/O.  If many sources are measured at the same set of sampling times and
have proportional measurement errors (\eg, an imaging instrument returning
to the same Field Of View many times) then the same $^{f}\alpha^{-1}$ matrix may be
reused on the $^{f}\beta$ vectors of all sources.  Additional speed-ups are also possible.

However, before implementing such optimizations, users should balance
the potential efficiency gains against the cost of the modification and determine whether the
reference implementation is already fast enough for their purposes.

As a test case, the reference implementation was applied to the 
Hipparcos Epoch Photometry Catalogue (HEP) \citep{1997hipp.conf...19L}.

Although the primary mission of Hipparcos was to provide high accuracy
positional astrometry, it also produced a large easily-available
high-quality photometry dataset.  The processed data includes
'Variability Annex 1' (hereafter VA1), listing variables with periods derived from
the the Hipparcos data, or periods from the previous literature
consistent with the Hipparcos data.
The HEP was later reprocessed by several groups, such as \citet{Koen},
to discover additional periodic variables.

The complete dataset of $>10^5$ stars, averaging $>10^2$ measurements each, 
spanning $>10^3$ days, takes $<10$ hours to search to frequencies up to
$(2 hours)^{-1}$ and $H=3$ on a standard desktop computer (a 2006 Apple Mac 
Pro with 2$\times$2 Intel cores at 2.66 GHz).  This is a factor of 2-3
times slower than the `Fast Computation of the Lomb Periodogram' code in 
{\it{Numerical Recipes}} \S 13.9, with comparable parameters for the search space,
using the same computer and compiler.  Profiling the code during
execution finds that the bulk of the CPU time is split roughly evenly between
the FFT and the linear regression stages.

Of the 118218 stars in the HEP, 115375 had sufficient high quality data to be processed 
by the reference implementation.  (Most of the others had interfering objects in
the field of view.)  Of these, 2275 had periods listed in VA1, $P_{VA1}$,
and had at least 50 measurements with no quality flags set
spanning more than $3 \times P_{VA1}$.

In the subset of 2275 VA1 stars, 2066 (88.5\%) had calculated periods that either agreed with
(50.0\%) or were harmonically related to $P_{VA1}$.  In descending order
of incidence, these harmonic relations were $2 \times P_{VA1}$ (15.7\%);
$1/2 \times P_{VA1}$ (12.8\%); $3  \times P_{VA1}$ (9.5\%); and
$3/2 \times P_{VA1}$ (0.6\%).  The other 262 stars (11.5\%)
had calculated periods that were unrelated to the HEP value.

The reference implementation has the ability to detrend the
the data with a polynomial fit (to search for periodic variation
on top of a slow irregular variation).  However applying this to the HEP
data, removing variation out to $t^3$, decreased the agreement with the VA1 
periods, increasing the number of harmonically unrelated periods from
11.5\% to 16.9\%, and decreasing the number at $P_{VA1}$ from
50.0\% to 42.4\%.  Koen \& Eyer recommend that if analysis with and
without such detrending find different periods, then the periods should
be considered unreliable.

\begin{figure}
\epsscale{.90}
\plotone{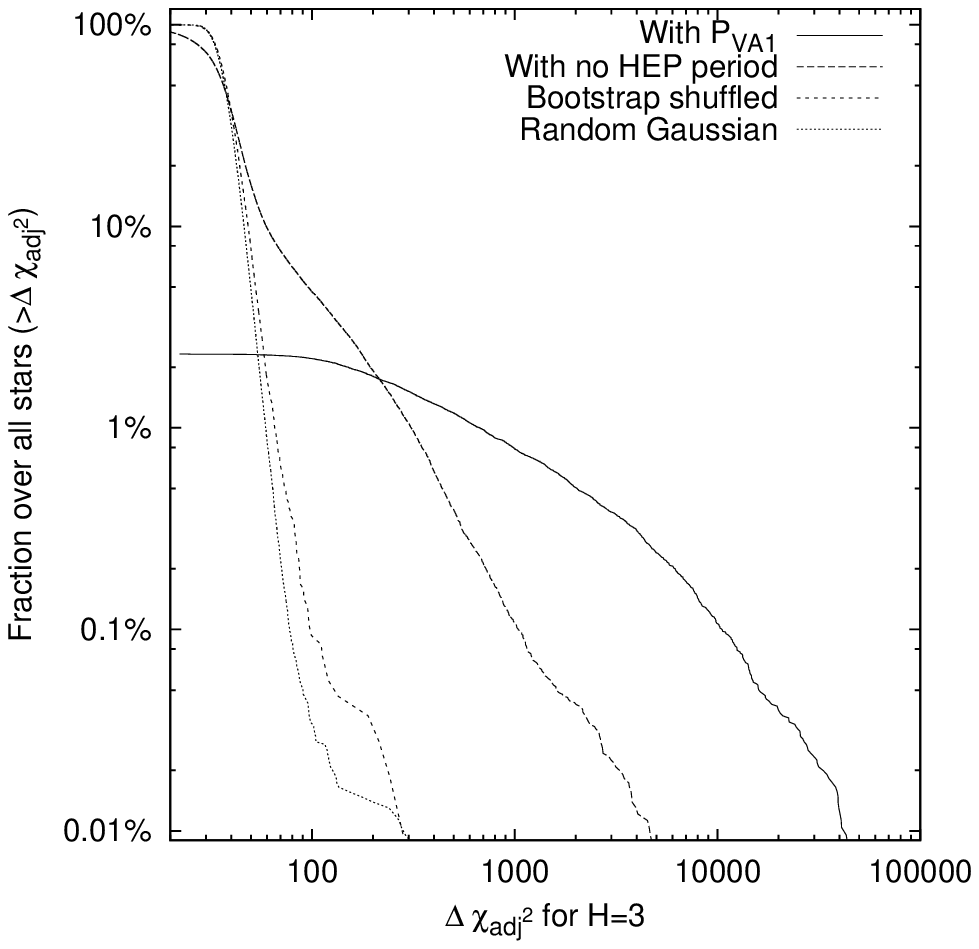}
\caption{Cumulative distribution of \dca2 for HEP stars with known periods from VA1 (solid line) and those without VA1 periods (dashed line) as a fraction of all HEP stars.  To test the false positive rate, the analysis was repeated on two randomized samples of the data.  In the {\it Bootstrap shuffle} (short-dashed line), the data for each star was shuffled so that each measurement value and its error estimate, $x_i \pm \sigma_i$, was randomly assigned to a measurement time $t_i$.   For the {\it Random Gaussian} (dotted line), the value at each $t_i$ was replaced by random variables with constant mean and standard deviation $C \pm \sigma$.\label{figfalsepos}}
\end{figure}

Figure~\ref{figfalsepos} shows the cumulative distribution of \dca2 for the HEP data
and for randomized data.  This demonstrates a well-behaved false-positive
rate when observing constant sources.  Excluding the VA1 stars (representing 2\% of the
population) a further $\sim$10\% of the HEP stars are clearly distinguishable from
randomized data by \fc2 algorithm.  This indicates variability in these stars, although
they do not have to be strictly periodic as long as there is significant power at some
period.

\begin{figure}
\plotone{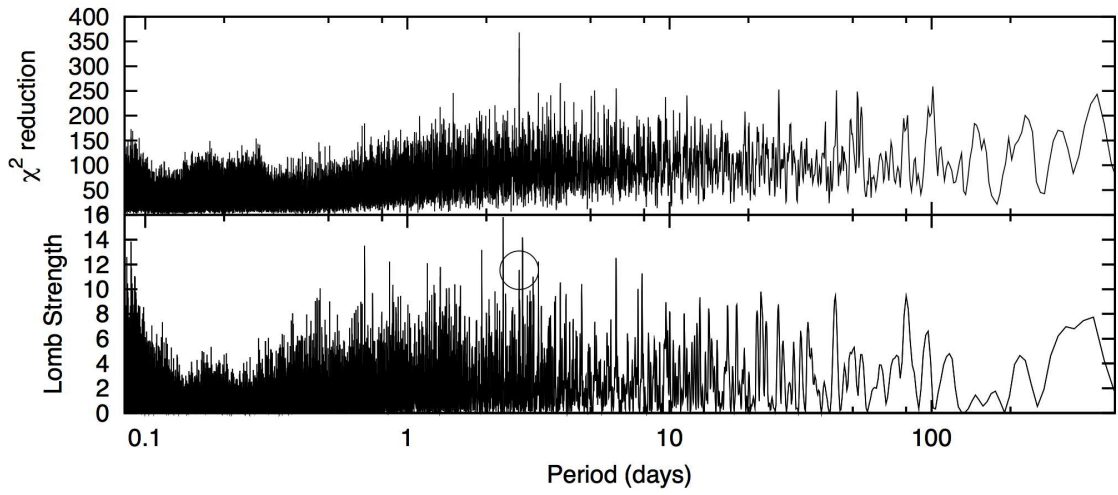}
\caption{Strength of the signal as a function of period for HIP 69358 for \fc2 (top) and Lomb (bottom) algorithms.  The Lomb peak corresponding to the \fc2 peak is circled.\label{figperiodogram}}
\end{figure}

Figure \ref{figperiodogram} shows a comparison of the results of the \fc2 and Lomb
algorithms for an example star, HIP 69358, which is listed as an
unsolved variable in the Hipparcos catalog.   A single strong peak at
$P_\mfc2 =  2.67098 d$ is consistent with the period of $2.67096 d$ 
found by \citet{Otero}.  However, the Lomb algorithm finds 12 peaks 
with higher strength than the one at that period.  As seen from the folded
light curves at the strongest $P_{Lomb}$ and $P_\mfc2$,
\fc2 found the characteristic light curve of an eclipsing binary
while Lomb found a noise peak.

Cases where \fc2 does not find the fundamental period are useful for
examining the limitations of this algorithm.  Some examples of this
are presented in Figure \ref{figlightcurves} and discussed in its caption.

\section{Conclusion}

The Fast \chisquared technique is a statistically efficient, statistically
valid method of searching for periodicity in data that may have
irregular sampling and non-uniform standard errors.

It is sensitive to power in the harmonics above the fundamental
frequency, to any arbitrary order.

It is computationally efficient, and can be composed largely
of standard FFT and linear algebra routines that are
commonly available in highly-optimized libraries.

A reference implementation is available 
and can be easily applied to your data set.

{\it Facilities:} \facility{HIPPARCOS}

\begin{figure}
\plotone{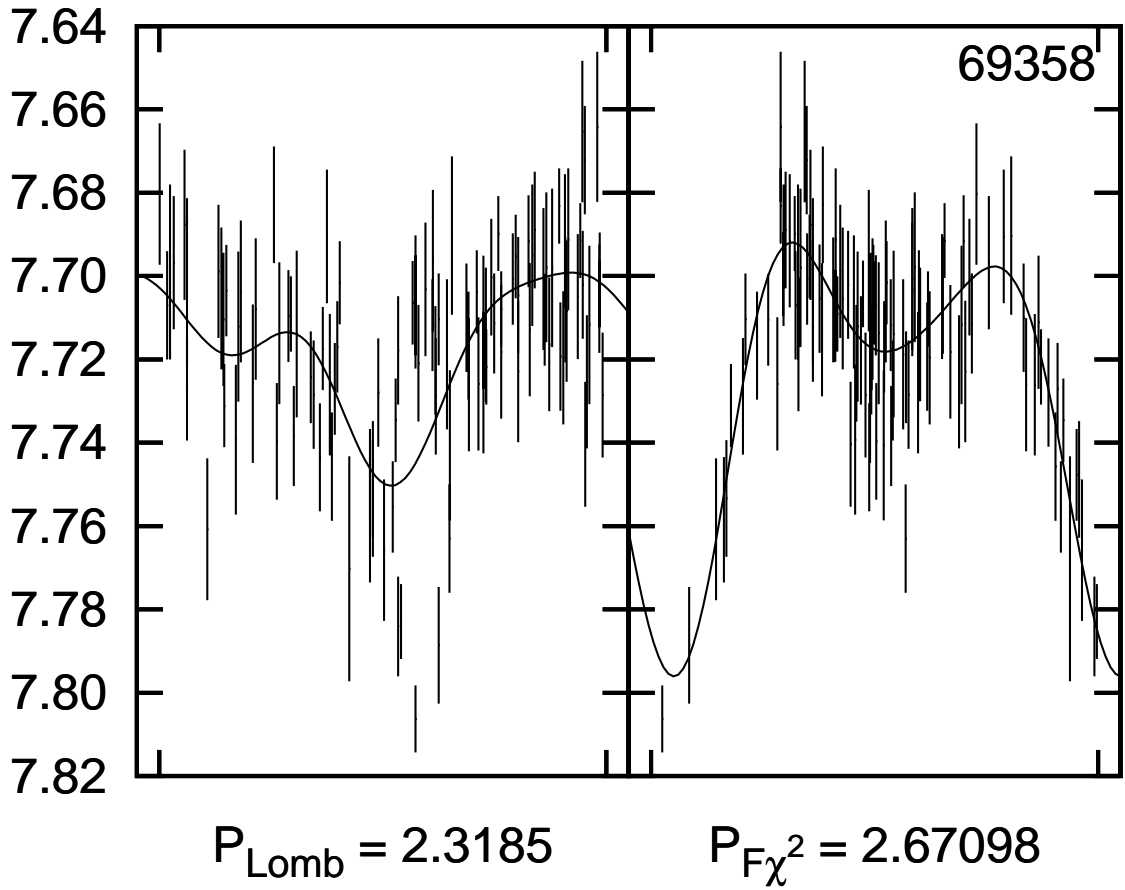}
\caption{HEP data for HIP 69358 folded at the periods found by Lomb (left) and \fc2 (right) algorithms.
\label{figlombcurve}}
\end{figure}

{
\begin{figure}
\caption{
Light curves of various stars folded at $P_{VA1}$ and $P_\mfc2$
to demonstrate some of the ways that \fc2 can find a period other than the 
fundamental of the physical system.  Star classification for these stars comes from the
Simbad Database, operated at CDS, Strasbourg, France
(
{\bf a)} {\it HIP~109303 == AR~Lac} is an RS~CVn eclipsing binary
with $P_{VA1} = 1.98318 d$ which \fc2 best-fits at half the period.  
Due to the narrow minima, the $H=3$ Fourier expansion at the fundamental,
\{$1f$, $2f$, $3f$\}, provides a worse fit than the fit at half the period \{$2f$, $4f$, $6f$\}.
However, the different depths of the two minima are clearly distinguishable,
showing unambiguously that $P_{VA1}$ is the true orbital period.
{\bf b)} {\it HIP~10701 == AD~Ari} has $P_{VA1} = 0.269862 d$.
However, \fc2 finds a period of twice that, which appears to
give a better fit to the data.  This is a $\delta$~Sct type
variable, which are characterized by multi-modal oscillations,
so power at multiple frequencies is to be expected.
{\bf c)}{\it HIP~59678 == DL~Cru} is an $\alpha$ Cyg variable.
These stars have complex oscillations, so different cycles
at the primary fundamental may have different apparent amplitudes.
The \fc2 algorithm to $H=3$, for this data set, found a lowest \chisquared at triple
the fundamental period.  This splits the cycles during the
observation into three sets, each of which can have a
quasi-independent average amplitude adjusted by the Fourier components,
allowing a slight additional \chisquared reduction through noise-fitting.
{\bf d)}{\it HIP~115647 == DP~Gru} is an Algol-type eclipsing variable
with narrow minima.  A fit to the best $H=3$ Fourier expansion is
not a good model for the shape of this lightcurve, and so the \fc2
would not be very effective in finding this period unless $H$ is increased.
However, the HEP data samples only 3 minima, spread over 145 orbits,
and so may not be sufficiently constraining to uniquely determine
a period, regardless of the algorithm used.
{\bf e)}{\it HIP~98546 == V1711~Sgr} is a W~Vir type variable.
$P_{VA1} = 15.052 d$, but \fc2 gives $10.566 d$.
The \fc2 fit is better in the formal sense of having a lower
\chisquared with the line passing closer to the data points, but it does
not look like a W Vir light curve.  To add to the confusion,
the General Catalog Of Variable Stars (
$P = 28.556 d$, but this period is not apparent
in the Hipparcos data.
{\bf f)}{\it HIP~12557 == W~Tri}, a semi-regular pulsating star, has 
$P_{VA1} = 108 d$, which is the third-best value found by \fc2 (after
 $17.99 d$ and $22.68 d$).
\label{figlightcurves}}
\end{figure}
\begin{figure}
{\plotone{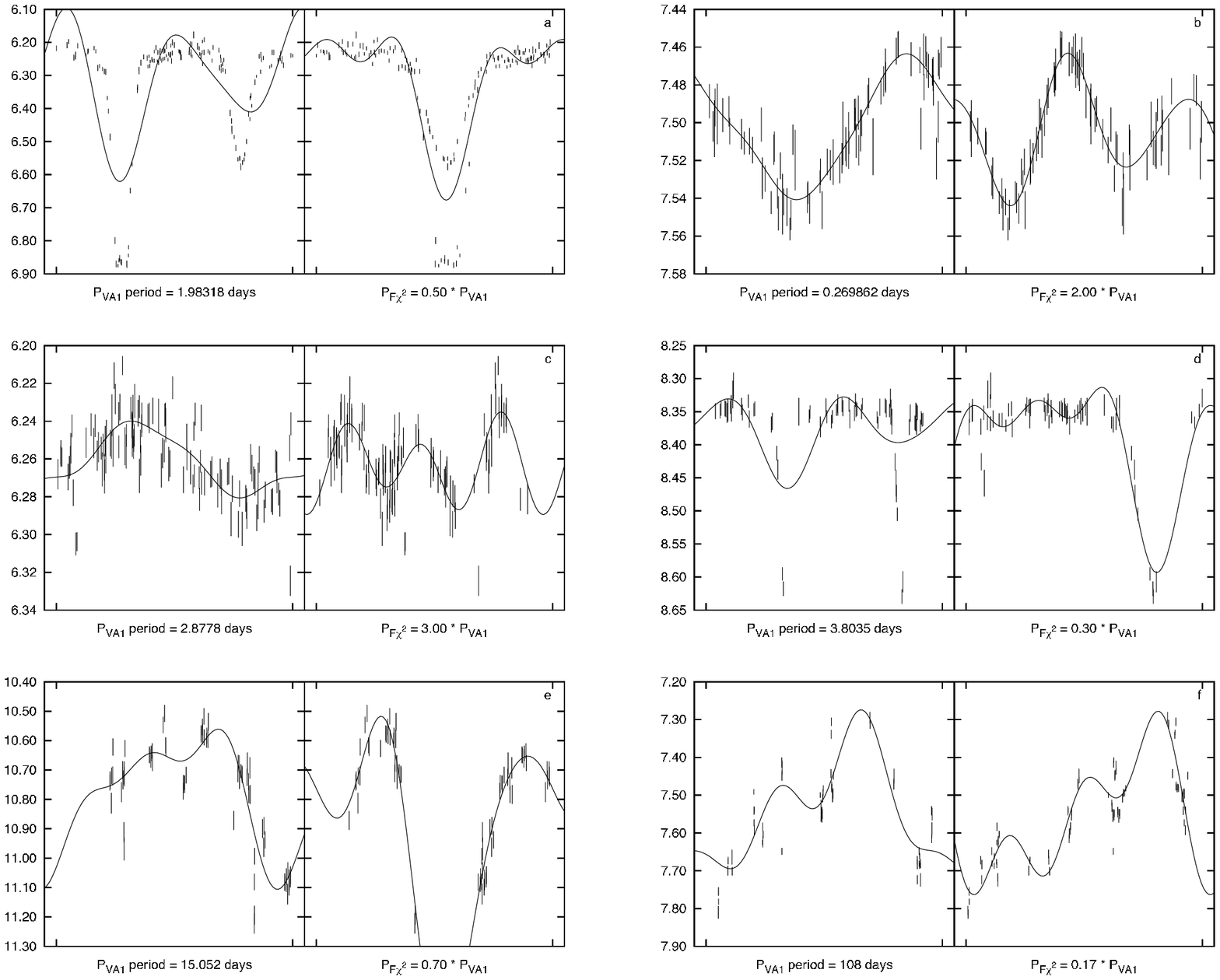}}
\end{figure}
}

\end{document}